\documentclass[prl,twocolumn,showpacs,preprintnumbers,nofootinbib]{revtex4-1}
\usepackage{graphicx}
\usepackage{dcolumn}
\usepackage{bm,amsfonts,amsthm,amsmath,amssymb}
\usepackage[]{epsfig,graphics}
\usepackage{comment}
\usepackage[T1]{fontenc}
\usepackage{lipsum}
\usepackage[utf8]{inputenc}
\usepackage{ulem}
\usepackage{multirow}
\usepackage{color}
\usepackage{lastpage}
\usepackage{enumerate}
\usepackage{subfigure}
\usepackage{wasysym}
\usepackage{hyperref}
\usepackage{graphicx}
\usepackage{caption}

\hypersetup{
    bookmarks=true,         
    unicode=false,          
    pdftoolbar=true,        
    pdfmenubar=true,        
    pdffitwindow=false,     
    pdfstartview={FitH},    
    pdftitle={My title},    
    pdfauthor={Author},     
    pdfsubject={Subject},   
    pdfcreator={Creator},   
    pdfproducer={Producer}, 
    pdfkeywords={keyword1} {key2} {key3}, 
    pdfnewwindow=true,      
    colorlinks=false,       
    linkcolor=red,          
    citecolor=green,        
    filecolor=magenta,      
    urlcolor=cyan           
}

\newenvironment{itemize*}
  {\begin{itemize}
    \setlength{\itemsep}{0pt}
    \setlength{\parskip}{0pt}}
  {\end{itemize}}

\newenvironment{enumerate*}
  {\begin{enumerate}
    \setlength{\itemsep}{0pt}
    \setlength{\parskip}{0pt}}
  {\end{enumerate}}

\newenvironment{description*}
  {\begin{description}
    \setlength{\itemsep}{0pt}
    \setlength{\parskip}{0pt}}
  {\end{description}}

\def\ben{\begin{enumerate*}}
\def\een{\end{enumerate*}}
\def\bi{\begin{itemize*}}
\def\ei{\end{itemize*}}
\def\bd{\begin{description*}}
\def\ed{\end{description*}}
\def\be{\begin{equation}}
\def\ee{\end{equation}}
\def\bea{\begin{eqnarray}}
\def\eea{\end{eqnarray}}
\def\bfl{\begin{flushleft}}
\def\efl{\end{flushleft}}

\newcommand{\frw}{{\mbox{\tiny FRW}}}
\newcommand{\osc}{{\mbox{\tiny osc}}}

\newcommand{\gsim}{\lower.7ex\hbox{$\;\stackrel{\textstyle>}{\sim}\;$}}
\newcommand{\lsim}{\lower.7ex\hbox{$\;\stackrel{\textstyle<}{\sim}\;$}}

\newcommand{\beq}{\begin{equation}}
\newcommand{\eeq}{\end{equation}}

\begin{document}

\title{A Model Independent Approach to (p)Reheating}
\author{Ogan \"Ozsoy}
\author{Gizem Sengor}
\author{Kuver Sinha}
\author{Scott Watson}
\affiliation{Department of Physics, Syracuse University, Syracuse, NY 13244, USA}
\date{\today}

\begin{abstract}
In this note we propose a model independent framework for inflationary (p)reheating.  Our approach is analogous to the Effective Field Theory of Inflation, however here the inflaton oscillations provide an additional source of (discrete) symmetry breaking.  Using the Goldstone field that non-linearly realizes time diffeormorphism invariance we construct a model independent action for both the inflaton and reheating sectors.
Utilizing the hierarchy of scales present during the reheating process we are able to recover known results in the literature in a simpler fashion, including the presence of oscillations in the primordial power spectrum.  
We also construct a class of models where the shift symmetry of the inflaton is preserved during reheating, which helps alleviate past criticisms of (p)reheating in models of Natural Inflation.  Extensions of our framework suggest the possibility of analytically investigating non-linear effects (such as rescattering and back-reaction) during thermalization without resorting to lattice methods. 
By construction, the EFT relates the strength of many of these interactions to other operators in the theory, including those responsible for the efficiency of (p)reheating.
We conclude with a discussion of the limitations and challenges for our approach.
\end{abstract}
\maketitle
The Effective Field Theory (EFT) of Inflation \cite{Creminelli:2006xe,Cheung:2007st,Senatore:2010wk} and generalizations to dark energy \cite{Creminelli:2008wc,Bloomfield:2012ff,Gubitosi:2012hu} and structure formation \cite{Baumann:2010tm} are based on the idea that there is a physical clock corresponding to the Goldstone boson that non-linearly realizes the spontaneously broken time diffeomorphism (diff)
invariance of the background.  In unitary gauge -- where the clock is homogeneous -- the matter perturbations are encoded within the metric, i.e. the would-be Goldstone bosons are 'eaten' by the metric since  gravity is a gauge theory.   

In this paper we would like to extend the EFT approach to include inflationary (p)reheating.  
There are many challenges.
In particular, during reheating the inflaton undergoes oscillations while quanta of reheating fields are produced in a process that can be quite involved and highly model-dependent. 
One of our goals in this paper is to alleviate some of this model dependence through the EFT approach and connect inflation with the late-time universe.
However, it is not obvious that the Goldstone language is useful in the presence of oscillations -- {\it Is the clock still monotonic?}  Moreover, as degrees of freedom of the inflaton are converted into reheat fields the composition of the Goldstone would seem to change if we consider that the matter driving the expansion (thus spontaneously breaking time diffs) changes from inflaton oscillations initially into reheating fields as time progresses.  In the remainder of this note we will address these issues, construct an EFT for reheating and present a few checks and applications of our approach\footnote{A detailed analysis will appear shortly in \cite{toappear}}.

\section{Reheating, Broken Symmetries and a Hierarchy of Scales}
Current observations require that if inflation took place it must have eventually ended leading to a hot and thermal universe by the time of Big Bang Nucleosynthesis.
The process by which the inflaton's energy is transferred into other particles is known as reheating.
This process can occur perturbatively \cite{Abbott:1982hn,Albrecht:1982mp,Dolgov:1982th}, or non-perturbatively in a process known as preheating \cite{Dolgov:1989us,Traschen:1990sw,Kofman:1997yn}. 
Existing investigations of (p)reheating typically rely on choosing a particular potential and then examining the choice of parameters that leads to a successful model.  However, one property that most models have in common is that particle production results during background oscillations of the inflaton field $\phi_0(t)$.

A key observation of our approach is that as long as the inflaton dominates the expansion the background will evolve as
\be \label{H_osc}
H(t) = H_{\frw}(t) + H_{\osc}(t) P(\omega t),
\ee
where the first term dominates and implies an adiabatically evolving, monotonically decreasing contribution to the expansion rate and the second term leads to an oscillatory correction which is sub-leading.
That is, $H_\frw > H_{\osc} P(\omega t)$ with $P(\omega t)$ a quasi-periodic function.
This evolution of the background spontaneously breaks time diffs $t \rightarrow t + \xi^0(t,\vec{x})$ first to a discrete symmetry\footnote{Discrete symmetry breaking {\it during inflation}
was considered in \cite{Behbahani:2011it}.} and then completely.
That is, if we probe the background at energies $E \gg \omega \gg H(t)$, time diffs are realized as a symmetry. 
This remains true until we consider energies comparable to $\omega$. At such energies
$H_{\frw}$ and $H_{\osc}$ will remain nearly constant preserving time diffs,
but the symmetry will be broken by  $P(\omega t)$ to a discrete symmetry $t \rightarrow t + 2\pi \omega^{-1}$. 
At lower energies both 
$H_{\frw}$ and $H_{\osc}$ will also evolve breaking the discrete symmetry.
This symmetry breaking pattern is a natural consequence of the hierarchy of scales that appears in reheating, i.e. high energy (small wavelength) modes probe inflaton oscillations $E / \omega$, whereas low energy (large wavelength) probes capture the expansion of the background $E / H_\frw$ and we have $H_\frw / \omega \ll 1$ during reheating.
We will use these facts below to construct the EFT of reheating in terms of the Goldstone boson that non-linearly realizes the broken time diffs.

Before proceeding we would like to be a little less abstract by what is meant by \eqref{H_osc}.
Consider a simple reheating model where the inflaton oscillates in a potential $V \simeq m^2 \phi_0^2$.
In this case we can solve for the background evolution 
and one finds\footnote{Here we present the growing mode solution and neglect an irrelevant decaying mode.} \cite{Mukhanov:2005sc}
\be \label{s1}
H(t)=H_m - \frac{3H_m^2}{4m} \sin (2mt+\delta) + \ldots, 
\ee
where $H_m = 2/(3t)$ is the Hubble rate in a matter dominated universe with scale factor $a(t) \sim t^{2/3}$, $\delta$ is a constant phase,
and dots represent terms suppressed by higher powers of $H_m/m$.
We see that \eqref{s1} is of the form of \eqref{H_osc} corresponding to a matter dominated universe corrected by oscillations suppressed by powers of $H_m/m$.
At energies comparable to the mass of the inflaton we have that the inflaton oscillations break the time diffs, whereas for energies $H \lesssim E \ll m$ the matter dominated expansion 
is primarily responsible for the breaking.  This is another way of stating the familiar fact that on scales comparable to the Hubble radius reheating with a massive inflaton oscillating in a quadratic potential looks like a matter dominated universe, whereas on small scales one can treat the particle production as a local process and in many cases neglect the presence of gravity.

In general the potential can be more complicated and the large scale evolution can depart from matter domination.  In particular, for an inflaton oscillating in a potential with leading term $V \sim \phi_0(t)^n$ the frequency
of oscillations will be $\omega \sim 1/(\int d\phi_0 (V_0 - V(\phi_0))^{-1/2} )$ where $V_0$ is the initial value of the potential and the frequency can depend on the initial amplitude of the inflaton \cite{Johnson:2008se}.  It is only for the case of harmonic motion that $\omega$ is constant.
For this general potential the leading term in \eqref{H_osc} is then given by $H_\frw = H_0 a^{-3n/(n+2)}$, which only corresponds to matter domination for $n=2$. 
At this point reheating looks rather model dependent. 
However our approach will remove the model dependence by focusing on a few properties shared by all such models. Firstly, the hierarchy $\omega >> H$ is preserved by all reasonable choices of potential.  Secondly, all we need to construct the EFT is that the background evolution spontaneously breaks time diff invariance.  As stressed by the authors of \cite{Cheung:2007st}, the background itself is not an observable and instead it is the perturbations about the background that are observable.  Noting these facts, we now proceed to construct our EFT for inflationary reheating.

\section{Inflaton decay, Reheating, and the Role of the Goldstone}
For successful models of reheating, the energy density will evolve from inflaton oscillations initially to a relativistic bath of particles.
This means that the fields responsible for spontaneously breaking the time diff invariance will change.
In \cite{Creminelli:2006xe} it was argued that the Goldstone approach goes beyond quasi-de Sitter (inflationary) spacetime and holds for any FRW universe
as given by \eqref{H_osc}.
Indeed, this holds for any number of matter fields $\phi_m(t)$ contributing to the background energy density with perturbation
\be
\delta \phi_m(t,\vec{x})= \phi_m\left(t+\pi(t,\vec{x})\right) - \phi_m(t).
\ee
This perturbation corresponds to a common, local shift in time for all the fields.   In the long wavelength limit this corresponds to the {\it adiabatic mode}, which Weinberg demonstrated obeys a conservation theorem regardless of the matter content of the universe \cite{Weinberg:2003sw}.  The field $\pi(x)$ is the desired Goldstone mode and we will use it to construct an EFT of reheating below.  
In addition to this mode, isocurvature modes may also be present -- especially given couplings of the inflaton to the reheat fields. Weinberg addressed this concern in \cite{Weinberg:2004kr} and demonstrated that for single field (single clock) inflation as the inflaton decays into reheat fields any source of non-adiabaticity would decay sufficiently fast to preserve the dominance of the adiabatic mode\footnote{This conclusion remains true even for multi-field inflation as long as local, thermal equilibrium with no non-zero conserved quantities occurs sometime following inflation \cite{Weinberg:2004kf}.}.   Given this, we will use $\pi(t,\vec{x})$ to construct our theory of reheating, noting that as inflatons are converted to reheat fields we are simply making use of the adiabatic mode description.   

The procedure for constructing the EFT follows analogously to that for inflation \cite{Cheung:2007st}.
Working in unitary gauge we construct the EFT of fluctuations for reheating in the gravitational and inflationary sectors as\footnote{We work in reduced Planck units $m_p=1/\sqrt{8 \pi G}=2.44 \times 10^{18}$ GeV with $\hbar=c=1$ and with a mostly plus $(-,+,+,+)$ sign convention for the metric.}
{\small
\bea 
S&=&\int d^4x \sqrt{-g}\, \left[ \frac{1}{2} m_p^2 R + m_p^2 \dot{H}g^{00} - m_p^2 \left(3H^2+\dot{H} \right) \right. \nonumber \\
&+&\left. \frac{M_2^4(t)}{2!} \left( \delta g^{00} \right)^2 +   \frac{M_3^4(t)}{3!} \left( \delta g^{00} \right)^3 + \ldots \right],
\label{theaction}
\eea
}where $g^{00}=-1+\delta g^{00}$ and the dots represent terms higher order in fluctuations and derivatives.  
Just as in the inflationary case we now introduce the Goldstone $\pi$ which non-linearly realizes time diffs.  This forces non-trivial relations between the
operators in \eqref{theaction} -- e.g. modifications to the sound speed and some interactions are both fixed by $M_2$.  
We see that because of the symmetry breaking the sound speed during inflaton oscillations is not protected and in general $c_\pi \neq 1$. 

One of the utilities of the EFT approach is that it is often useful to take a decoupling limit where $\dot{H} \rightarrow 0$ and $m^2_p \rightarrow \infty$ while the combination remains fixed.
This limit makes more precise the usual assumption in (p)reheating that particle creation is a local process and one typically ignores contributions coming from gravitational terms\footnote{During preheating 
modes within a particular resonance band can redshift, but it was shown in \cite{Kofman:1997yn} that this effect can be neglected since the modes
that initially start in the dominant resonance band give the dominant contribution to production.} when calculating 
the number of particles produced \cite{Kofman:1997yn}.

For now we focus on operators fixed by tadpole cancelation and take $M_2=M_3=\ldots=0$.  In spatially flat gauge the quadratic action in the decoupling limit is
{\small
\be \label{decoupled_action}
S_2=\int d^4x \,a^3 m_p^2 \left[ -\dot{H}\left(\dot{\pi}^2-a^{-2}(\partial_i \pi )^2 \right) -3\dot{H}^2 \pi^2 \right],
\ee
}which by canceling the tadpoles has left us with coefficients fixed by the background evolution.
Introducing the canonical field ${\pi}_c =m_p (-2 \dot{H})^{1/2} {\pi}$ one can show that the term responsible for breaking the shift symmetry is due to the oscillatory behavior of the time-dependent potential of the inflaton (corresponding to an operator $\hat{{\cal O}}_\pi \sim V^{\prime \prime} \pi_c^2$) and does {\it not} come from mixing with gravity.
As in the EFT of Inflation the leading mixing with gravity scales as $E_{\mbox{\tiny mix}}=\epsilon^{1/2} H=\dot{H}^{1/2}$, although a difference for us is given $V\sim \phi_0^n$ then $\epsilon=3n/(n+2)$ is typically an order one number.
The decoupling limit will be useful for probing scales with $E \gg E_{\mbox{\tiny mix}}$, but other times it is appropriate to include corrections coming from the mixing with gravity.
One useful aspect of \eqref{decoupled_action} is to study the stability of sub-horizon perturbations against collapse. Just as in studies of ghost condensation \cite{ArkaniHamed:2003uy}, including higher corrections to the EFT (e.g. $M_2 \neq 0$) could lead to new and consistent models for (p)reheating.

Depending on the question one is asking it may or may not be useful to take the decoupling limit.
For example, in understanding the behavior of modes that are re-entering the horizon during reheating it is useful to calculate the leading corrections to 
\eqref{decoupled_action} coming from the mixing with gravity. In that case, we consider modes between the two hierarchical scales $k/(am) \ll 1$ while $k/(aH) \gtrsim 1$.
The leading mixing term is
\be
\Delta S_2 = -\frac{1}{2} \int d^4x \, a^3 \left( \frac{2 \ddot{H}}{H}\right) \pi_c^2,
\ee
which we have written in terms of the canoncial field $\pi_c$.
One can show that this leading correction results in a 
growing, oscillatory contribution to the power spectrum.
For example, if we choose the particular case $V\sim m^2 \phi_0^2$, we find this correction leads to the main result of \cite{Easther:2010mr},
where those authors performed a full analysis (including all gravitational perturbations).

Part of the utility of our approach is that inflaton self-interactions will also be fixed by the symmetries. For (p)reheating this implies that if one is interested in interactions, which determine rescattering and backreaction effects, the coefficients for these terms that appear in the action will also be fixed by the same symmetries. We postpone further discussion of this for the future and instead turn to 
the issue of particle production of the reheat field.

\section{Reheating Fields and Coupling to the Inflaton}
Given the EFT description above, we now couple the inflationary sector to an additional reheat field $\sigma(t,\vec{x})$.
We will be interested in the production of $\sigma(t,\vec{x})$ particles resulting from the oscillations of the background inflaton field $\phi_0(t)$.
In unitary gauge, the production of particles by the background will result from operators\footnote{We do not require the reheat field to be shift symmetric.} $f(t) \hat{{\cal O}}_n(\sigma)$.
At the quadratic level this gives
{\small
\bea 
S^{(2)}_{\sigma} &=&\int d^4x\sqrt{-g}\left[-\frac{\alpha_1(t)}{2}g^{\mu\nu}\partial_\mu\sigma\partial_\nu\sigma \right. \nonumber \\
&+&\left.\frac{\alpha_2(t)}{2}(\partial^o\sigma)^2-\frac{\alpha_3(t)}{2}\sigma^2+\alpha_4(t)\sigma\partial^o\sigma\right], \;\;
\label{sigma_action}
\eea
}where we see that the broken time diffs allow for a non-trivial sound speed $c_\sigma^2=\alpha_1/(\alpha_1+\alpha_2)$.
The action \eqref{sigma_action} already accounts for many existing models in the literature.  For example, preheating with $V \sim g^2 \phi_0^2 \sigma^2$ corresponds to 
$\alpha_1=1,\alpha_2=\alpha_4=0$ and $\alpha_3=g^2 \phi_0(t)^2$.  Whereas, if we require the inflaton to remain shift symmetric throughout reheating, 
as one might anticipate in models of Natural Inflation, then we consider interactions of the form $(\partial_\mu \phi_0)^2 \sigma^2/\Lambda^2$,
where $\Lambda$ is the cutoff for the {\it background}.
Our approach captures this model by now choosing $\alpha_3=2\dot{\phi}_0(t)^2/\Lambda^2$. 
It has been found that preheating in models that preserve an inflaton shift symmetry is not efficient \cite{ArmendarizPicon:2007iv}.
One reason for this is that naively we assume that the energy of the fields can not exceed the cutoff $\Lambda$.
However, an advantage of our EFT approach is that the parameters, such as
$\alpha_3$, can be completely non-linear and their origin is irrelevant since the background itself is not physically observable.
This is analogous to the EFT of Inflation, where noting that the background is not an observable the authors assume, a priori, a quasi-de Sitter background and then study the EFT
of fluctuations about that background.

Although the parameters are not directly observable, we can constrain the EFT parameters in several ways.  
Just as in the case of the inflaton above, avoiding instabilities will place constraints.
Moreover, we must require that the coefficients violate adiabaticity so that particles are produced --
this implies that $\dot{\alpha}_3 / \alpha^2_3 \gg1$. 
When adiabaticity fails $\sigma$-quanta will be produced.  A calculation of the precise number of particles that results from the violation 
would require a detailed investigation and would seem to still be model dependent.
However, an advantage of our approach is that these same coefficients will enter into interaction terms with their form dictated by non-linearly realizing time diffs.
Interactions between inflaton particles ($\delta \phi \sim \pi$) and the reheat particles ($\sigma$) will have two origins in the EFT.
Firstly, upon reintroducing the Goldstone field via $t \rightarrow t+\pi$ the coefficients in \eqref{sigma_action} will generate interactions.  Secondly,
there will be mixing terms of the form $\sim (\delta g^{00})^m \hat{{\cal O}}_n(\sigma)$. As an example of the first type there is an operator $\sim \dot{\alpha}_3 \pi \sigma^2$ 
and so we see when adiabaticity is violated $\dot{\alpha}_3 \gg  \alpha_3^2$ this interaction would become important.
Details will appear in \cite{toappear}, here we just want to emphasize that the symmetries will enforce connections like this and so 
our approach connects successful particle production (violations of adiabaticity) with the interactions that are important for understanding
particle back-reaction, rescattering, and the details of thermalization.

\section{Remarks and Conclusions} 
In this note we have outlined a program to establish a model independent approach to (p)reheating.
We have seen that requiring the Goldstone to non-linearly realize the broken time diffeomorophisms implies important connections
between the background dependent parameters of the EFT. In particular, we have seen that the amount of particle production and the sound speed
is related to the strength of interactions which are important for understanding the duration of (p)reheating and the thermalization process.  

There are some challenges for our approach.  We have assumed initially that the reheat field does not substantially contribute to the energy density.
This assumption is certainly justified during the first phases of preheating \cite{Kofman:1997yn}, however as particles continue to be produced they may influence the duration of reheating.
In the case of the EFT of inflation this can cause an ambiguity in the (clock) Goldstone description \cite{LopezNacir:2011kk}.  Here we have argued that by tracking the adiabatic mode and neglecting 
any isocurvature we can avoid this problem, but this issue requires a more careful investigation.  Another issue is that unlike the EFT of inflation where primordial non-gaussianity (or lack of it) was an important observation for restricting the parameters of the EFT, such observations seem rather irrelevant\footnote{A possible exception would be in models where the couplings both during inflation and after remain the same.  Then non-Gaussianity constraints could be used to restrict the EFT of reheating.} for (p)reheating.  Instead, aside from the possibility of gravity wave signatures, our main application is developing a framework to analytically understand the thermalization process\footnote{We thank M. Peloso for discussions.}. As we have seen above, the form and strength of these interactions are restricted by the symmetries. This, along with the model independence that results from our treatment of the background suggests this is a promising approach to further develop. 

\section*{Acknowledgements}
We thank Peter Adshead, Rouzbeh Allahverdi, Mustafa Amin, Daniel Baumann, Sera Cremonini and Marco Peloso for useful discussions.  This work was supported in part by NASA Astrophysics Theory Grant NNH12ZDA001N, DOE grant DE-FG02-85ER40237, National Science Foundation Grant No. PHYS-1066293 and the hospitality of the Aspen Center for Physics.

\bibliographystyle{apsrev4-1}

\begin{thebibliography}{24}%
\makeatletter
\providecommand \@ifxundefined [1]{%
 \@ifx{#1\undefined}
}%
\providecommand \@ifnum [1]{%
 \ifnum #1\expandafter \@firstoftwo
 \else \expandafter \@secondoftwo
 \fi
}%
\providecommand \@ifx [1]{%
 \ifx #1\expandafter \@firstoftwo
 \else \expandafter \@secondoftwo
 \fi
}%
\providecommand \natexlab [1]{#1}%
\providecommand \enquote  [1]{``#1''}%
\providecommand \bibnamefont  [1]{#1}%
\providecommand \bibfnamefont [1]{#1}%
\providecommand \citenamefont [1]{#1}%
\providecommand \href@noop [0]{\@secondoftwo}%
\providecommand \href [0]{\begingroup \@sanitize@url \@href}%
\providecommand \@href[1]{\@@startlink{#1}\@@href}%
\providecommand \@@href[1]{\endgroup#1\@@endlink}%
\providecommand \@sanitize@url [0]{\catcode `\\12\catcode `\$12\catcode
  `\&12\catcode `\#12\catcode `\^12\catcode `\_12\catcode `\%12\relax}%
\providecommand \@@startlink[1]{}%
\providecommand \@@endlink[0]{}%
\providecommand \url  [0]{\begingroup\@sanitize@url \@url }%
\providecommand \@url [1]{\endgroup\@href {#1}{\urlprefix }}%
\providecommand \urlprefix  [0]{URL }%
\providecommand \Eprint [0]{\href }%
\providecommand \doibase [0]{http://dx.doi.org/}%
\providecommand \selectlanguage [0]{\@gobble}%
\providecommand \bibinfo  [0]{\@secondoftwo}%
\providecommand \bibfield  [0]{\@secondoftwo}%
\providecommand \translation [1]{[#1]}%
\providecommand \BibitemOpen [0]{}%
\providecommand \bibitemStop [0]{}%
\providecommand \bibitemNoStop [0]{.\EOS\space}%
\providecommand \EOS [0]{\spacefactor3000\relax}%
\providecommand \BibitemShut  [1]{\csname bibitem#1\endcsname}%
\let\auto@bib@innerbib\@empty
\bibitem [{\citenamefont {Creminelli}\ \emph {et~al.}(2006)\citenamefont
  {Creminelli}, \citenamefont {Luty}, \citenamefont {Nicolis},\ and\
  \citenamefont {Senatore}}]{Creminelli:2006xe}%
  \BibitemOpen
  \bibfield  {author} {\bibinfo {author} {\bibfnamefont {P.}~\bibnamefont
  {Creminelli}}, \bibinfo {author} {\bibfnamefont {M.~A.}\ \bibnamefont
  {Luty}}, \bibinfo {author} {\bibfnamefont {A.}~\bibnamefont {Nicolis}}, \
  and\ \bibinfo {author} {\bibfnamefont {L.}~\bibnamefont {Senatore}},\ }\href
  {\doibase 10.1088/1126-6708/2006/12/080} {\bibfield  {journal} {\bibinfo
  {journal} {JHEP}\ }\textbf {\bibinfo {volume} {0612}},\ \bibinfo {pages}
  {080} (\bibinfo {year} {2006})},\ \Eprint
  {http://arxiv.org/abs/hep-th/0606090} {arXiv:hep-th/0606090 [hep-th]}
  \BibitemShut {NoStop}%
\bibitem [{\citenamefont {Cheung}\ \emph {et~al.}(2008)\citenamefont {Cheung},
  \citenamefont {Creminelli}, \citenamefont {Fitzpatrick}, \citenamefont
  {Kaplan},\ and\ \citenamefont {Senatore}}]{Cheung:2007st}%
  \BibitemOpen
  \bibfield  {author} {\bibinfo {author} {\bibfnamefont {C.}~\bibnamefont
  {Cheung}}, \bibinfo {author} {\bibfnamefont {P.}~\bibnamefont {Creminelli}},
  \bibinfo {author} {\bibfnamefont {A.~L.}\ \bibnamefont {Fitzpatrick}},
  \bibinfo {author} {\bibfnamefont {J.}~\bibnamefont {Kaplan}}, \ and\ \bibinfo
  {author} {\bibfnamefont {L.}~\bibnamefont {Senatore}},\ }\href {\doibase
  10.1088/1126-6708/2008/03/014} {\bibfield  {journal} {\bibinfo  {journal}
  {JHEP}\ }\textbf {\bibinfo {volume} {0803}},\ \bibinfo {pages} {014}
  (\bibinfo {year} {2008})},\ \Eprint {http://arxiv.org/abs/0709.0293}
  {arXiv:0709.0293 [hep-th]} \BibitemShut {NoStop}%
\bibitem [{\citenamefont {Senatore}\ and\ \citenamefont
  {Zaldarriaga}(2010)}]{Senatore:2010wk}%
  \BibitemOpen
  \bibfield  {author} {\bibinfo {author} {\bibfnamefont {L.}~\bibnamefont
  {Senatore}}\ and\ \bibinfo {author} {\bibfnamefont {M.}~\bibnamefont
  {Zaldarriaga}},\ }\href@noop {} {\  (\bibinfo {year} {2010})},\ \Eprint
  {http://arxiv.org/abs/1009.2093} {arXiv:1009.2093 [hep-th]} \BibitemShut
  {NoStop}%
\bibitem [{\citenamefont {Creminelli}\ \emph {et~al.}(2009)\citenamefont
  {Creminelli}, \citenamefont {D'Amico}, \citenamefont {Norena},\ and\
  \citenamefont {Vernizzi}}]{Creminelli:2008wc}%
  \BibitemOpen
  \bibfield  {author} {\bibinfo {author} {\bibfnamefont {P.}~\bibnamefont
  {Creminelli}}, \bibinfo {author} {\bibfnamefont {G.}~\bibnamefont {D'Amico}},
  \bibinfo {author} {\bibfnamefont {J.}~\bibnamefont {Norena}}, \ and\ \bibinfo
  {author} {\bibfnamefont {F.}~\bibnamefont {Vernizzi}},\ }\href {\doibase
  10.1088/1475-7516/2009/02/018} {\bibfield  {journal} {\bibinfo  {journal}
  {JCAP}\ }\textbf {\bibinfo {volume} {0902}},\ \bibinfo {pages} {018}
  (\bibinfo {year} {2009})},\ \Eprint {http://arxiv.org/abs/0811.0827}
  {arXiv:0811.0827 [astro-ph]} \BibitemShut {NoStop}%
\bibitem [{\citenamefont {Bloomfield}\ \emph {et~al.}(2013)\citenamefont
  {Bloomfield}, \citenamefont {Flanagan}, \citenamefont {Park},\ and\
  \citenamefont {Watson}}]{Bloomfield:2012ff}%
  \BibitemOpen
  \bibfield  {author} {\bibinfo {author} {\bibfnamefont {J.~K.}\ \bibnamefont
  {Bloomfield}}, \bibinfo {author} {\bibfnamefont {E.~E.}\ \bibnamefont
  {Flanagan}}, \bibinfo {author} {\bibfnamefont {M.}~\bibnamefont {Park}}, \
  and\ \bibinfo {author} {\bibfnamefont {S.}~\bibnamefont {Watson}},\ }\href
  {\doibase 10.1088/1475-7516/2013/08/010} {\bibfield  {journal} {\bibinfo
  {journal} {JCAP}\ }\textbf {\bibinfo {volume} {1308}},\ \bibinfo {pages}
  {010} (\bibinfo {year} {2013})},\ \Eprint {http://arxiv.org/abs/1211.7054}
  {arXiv:1211.7054 [astro-ph.CO]} \BibitemShut {NoStop}%
\bibitem [{\citenamefont {Gubitosi}\ \emph {et~al.}(2013)\citenamefont
  {Gubitosi}, \citenamefont {Piazza},\ and\ \citenamefont
  {Vernizzi}}]{Gubitosi:2012hu}%
  \BibitemOpen
  \bibfield  {author} {\bibinfo {author} {\bibfnamefont {G.}~\bibnamefont
  {Gubitosi}}, \bibinfo {author} {\bibfnamefont {F.}~\bibnamefont {Piazza}}, \
  and\ \bibinfo {author} {\bibfnamefont {F.}~\bibnamefont {Vernizzi}},\ }\href
  {\doibase 10.1088/1475-7516/2013/02/032} {\bibfield  {journal} {\bibinfo
  {journal} {JCAP}\ }\textbf {\bibinfo {volume} {1302}},\ \bibinfo {pages}
  {032} (\bibinfo {year} {2013})},\ \bibinfo {note} {[JCAP1302,032(2013)]},\
  \Eprint {http://arxiv.org/abs/1210.0201} {arXiv:1210.0201 [hep-th]}
  \BibitemShut {NoStop}%
\bibitem [{\citenamefont {Baumann}\ \emph {et~al.}(2012)\citenamefont
  {Baumann}, \citenamefont {Nicolis}, \citenamefont {Senatore},\ and\
  \citenamefont {Zaldarriaga}}]{Baumann:2010tm}%
  \BibitemOpen
  \bibfield  {author} {\bibinfo {author} {\bibfnamefont {D.}~\bibnamefont
  {Baumann}}, \bibinfo {author} {\bibfnamefont {A.}~\bibnamefont {Nicolis}},
  \bibinfo {author} {\bibfnamefont {L.}~\bibnamefont {Senatore}}, \ and\
  \bibinfo {author} {\bibfnamefont {M.}~\bibnamefont {Zaldarriaga}},\ }\href
  {\doibase 10.1088/1475-7516/2012/07/051} {\bibfield  {journal} {\bibinfo
  {journal} {JCAP}\ }\textbf {\bibinfo {volume} {1207}},\ \bibinfo {pages}
  {051} (\bibinfo {year} {2012})},\ \Eprint {http://arxiv.org/abs/1004.2488}
  {arXiv:1004.2488 [astro-ph.CO]} \BibitemShut {NoStop}%
\bibitem [{\citenamefont {Ozsoy}\ \emph {et~al.}()\citenamefont {Ozsoy},
  \citenamefont {Sengor}, \citenamefont {Sinha},\ and\ \citenamefont
  {Watson}}]{toappear}%
  \BibitemOpen
  \bibfield  {author} {\bibinfo {author} {\bibfnamefont {O.}~\bibnamefont
  {Ozsoy}}, \bibinfo {author} {\bibfnamefont {G.}~\bibnamefont {Sengor}},
  \bibinfo {author} {\bibfnamefont {K.}~\bibnamefont {Sinha}}, \ and\ \bibinfo
  {author} {\bibfnamefont {S.}~\bibnamefont {Watson}},\ }\href@noop {} {\
  }\bibinfo {note} {To appear}\BibitemShut {NoStop}%
\bibitem [{\citenamefont {Abbott}\ \emph {et~al.}(1982)\citenamefont {Abbott},
  \citenamefont {Farhi},\ and\ \citenamefont {Wise}}]{Abbott:1982hn}%
  \BibitemOpen
  \bibfield  {author} {\bibinfo {author} {\bibfnamefont {L.~F.}\ \bibnamefont
  {Abbott}}, \bibinfo {author} {\bibfnamefont {E.}~\bibnamefont {Farhi}}, \
  and\ \bibinfo {author} {\bibfnamefont {M.~B.}\ \bibnamefont {Wise}},\ }\href
  {\doibase 10.1016/0370-2693(82)90867-X} {\bibfield  {journal} {\bibinfo
  {journal} {Phys. Lett.}\ }\textbf {\bibinfo {volume} {B117}},\ \bibinfo
  {pages} {29} (\bibinfo {year} {1982})}\BibitemShut {NoStop}%
\bibitem [{\citenamefont {Albrecht}\ \emph {et~al.}(1982)\citenamefont
  {Albrecht}, \citenamefont {Steinhardt}, \citenamefont {Turner},\ and\
  \citenamefont {Wilczek}}]{Albrecht:1982mp}%
  \BibitemOpen
  \bibfield  {author} {\bibinfo {author} {\bibfnamefont {A.}~\bibnamefont
  {Albrecht}}, \bibinfo {author} {\bibfnamefont {P.~J.}\ \bibnamefont
  {Steinhardt}}, \bibinfo {author} {\bibfnamefont {M.~S.}\ \bibnamefont
  {Turner}}, \ and\ \bibinfo {author} {\bibfnamefont {F.}~\bibnamefont
  {Wilczek}},\ }\href {\doibase 10.1103/PhysRevLett.48.1437} {\bibfield
  {journal} {\bibinfo  {journal} {Phys. Rev. Lett.}\ }\textbf {\bibinfo
  {volume} {48}},\ \bibinfo {pages} {1437} (\bibinfo {year}
  {1982})}\BibitemShut {NoStop}%
\bibitem [{\citenamefont {Dolgov}\ and\ \citenamefont
  {Linde}(1982)}]{Dolgov:1982th}%
  \BibitemOpen
  \bibfield  {author} {\bibinfo {author} {\bibfnamefont {A.~D.}\ \bibnamefont
  {Dolgov}}\ and\ \bibinfo {author} {\bibfnamefont {A.~D.}\ \bibnamefont
  {Linde}},\ }\href {\doibase 10.1016/0370-2693(82)90292-1} {\bibfield
  {journal} {\bibinfo  {journal} {Phys. Lett.}\ }\textbf {\bibinfo {volume}
  {B116}},\ \bibinfo {pages} {329} (\bibinfo {year} {1982})}\BibitemShut
  {NoStop}%
\bibitem [{\citenamefont {Dolgov}\ and\ \citenamefont
  {Kirilova}(1990)}]{Dolgov:1989us}%
  \BibitemOpen
  \bibfield  {author} {\bibinfo {author} {\bibfnamefont {A.~D.}\ \bibnamefont
  {Dolgov}}\ and\ \bibinfo {author} {\bibfnamefont {D.~P.}\ \bibnamefont
  {Kirilova}},\ }\href@noop {} {\bibfield  {journal} {\bibinfo  {journal} {Sov.
  J. Nucl. Phys.}\ }\textbf {\bibinfo {volume} {51}},\ \bibinfo {pages} {172}
  (\bibinfo {year} {1990})},\ \bibinfo {note} {[Yad.
  Fiz.51,273(1990)]}\BibitemShut {NoStop}%
\bibitem [{\citenamefont {Traschen}\ and\ \citenamefont
  {Brandenberger}(1990)}]{Traschen:1990sw}%
  \BibitemOpen
  \bibfield  {author} {\bibinfo {author} {\bibfnamefont {J.~H.}\ \bibnamefont
  {Traschen}}\ and\ \bibinfo {author} {\bibfnamefont {R.~H.}\ \bibnamefont
  {Brandenberger}},\ }\href {\doibase 10.1103/PhysRevD.42.2491} {\bibfield
  {journal} {\bibinfo  {journal} {Phys. Rev.}\ }\textbf {\bibinfo {volume}
  {D42}},\ \bibinfo {pages} {2491} (\bibinfo {year} {1990})}\BibitemShut
  {NoStop}%
\bibitem [{\citenamefont {Kofman}\ \emph {et~al.}(1997)\citenamefont {Kofman},
  \citenamefont {Linde},\ and\ \citenamefont {Starobinsky}}]{Kofman:1997yn}%
  \BibitemOpen
  \bibfield  {author} {\bibinfo {author} {\bibfnamefont {L.}~\bibnamefont
  {Kofman}}, \bibinfo {author} {\bibfnamefont {A.~D.}\ \bibnamefont {Linde}}, \
  and\ \bibinfo {author} {\bibfnamefont {A.~A.}\ \bibnamefont {Starobinsky}},\
  }\href {\doibase 10.1103/PhysRevD.56.3258} {\bibfield  {journal} {\bibinfo
  {journal} {Phys.Rev.}\ }\textbf {\bibinfo {volume} {D56}},\ \bibinfo {pages}
  {3258} (\bibinfo {year} {1997})},\ \Eprint
  {http://arxiv.org/abs/hep-ph/9704452} {arXiv:hep-ph/9704452 [hep-ph]}
  \BibitemShut {NoStop}%
\bibitem [{\citenamefont {Behbahani}\ \emph {et~al.}(2012)\citenamefont
  {Behbahani}, \citenamefont {Dymarsky}, \citenamefont {Mirbabayi},\ and\
  \citenamefont {Senatore}}]{Behbahani:2011it}%
  \BibitemOpen
  \bibfield  {author} {\bibinfo {author} {\bibfnamefont {S.~R.}\ \bibnamefont
  {Behbahani}}, \bibinfo {author} {\bibfnamefont {A.}~\bibnamefont {Dymarsky}},
  \bibinfo {author} {\bibfnamefont {M.}~\bibnamefont {Mirbabayi}}, \ and\
  \bibinfo {author} {\bibfnamefont {L.}~\bibnamefont {Senatore}},\ }\href
  {\doibase 10.1088/1475-7516/2012/12/036} {\bibfield  {journal} {\bibinfo
  {journal} {JCAP}\ }\textbf {\bibinfo {volume} {1212}},\ \bibinfo {pages}
  {036} (\bibinfo {year} {2012})},\ \Eprint {http://arxiv.org/abs/1111.3373}
  {arXiv:1111.3373 [hep-th]} \BibitemShut {NoStop}%
\bibitem [{\citenamefont {Mukhanov}(2005)}]{Mukhanov:2005sc}%
  \BibitemOpen
  \bibfield  {author} {\bibinfo {author} {\bibfnamefont {V.}~\bibnamefont
  {Mukhanov}},\ }\href
  {http://www-spires.fnal.gov/spires/find/books/www?cl=QB981.M89::2005} {\emph
  {\bibinfo {title} {{Physical Foundations of Cosmology}}}}\ (\bibinfo
  {publisher} {Cambridge University Press},\ \bibinfo {address} {Oxford},\
  \bibinfo {year} {2005})\BibitemShut {NoStop}%
\bibitem [{\citenamefont {Johnson}\ and\ \citenamefont
  {Kamionkowski}(2008)}]{Johnson:2008se}%
  \BibitemOpen
  \bibfield  {author} {\bibinfo {author} {\bibfnamefont {M.~C.}\ \bibnamefont
  {Johnson}}\ and\ \bibinfo {author} {\bibfnamefont {M.}~\bibnamefont
  {Kamionkowski}},\ }\href {\doibase 10.1103/PhysRevD.78.063010} {\bibfield
  {journal} {\bibinfo  {journal} {Phys. Rev.}\ }\textbf {\bibinfo {volume}
  {D78}},\ \bibinfo {pages} {063010} (\bibinfo {year} {2008})},\ \Eprint
  {http://arxiv.org/abs/0805.1748} {arXiv:0805.1748 [astro-ph]} \BibitemShut
  {NoStop}%
\bibitem [{\citenamefont {Weinberg}(2003)}]{Weinberg:2003sw}%
  \BibitemOpen
  \bibfield  {author} {\bibinfo {author} {\bibfnamefont {S.}~\bibnamefont
  {Weinberg}},\ }\href {\doibase 10.1103/PhysRevD.67.123504} {\bibfield
  {journal} {\bibinfo  {journal} {Phys.Rev.}\ }\textbf {\bibinfo {volume}
  {D67}},\ \bibinfo {pages} {123504} (\bibinfo {year} {2003})},\ \Eprint
  {http://arxiv.org/abs/astro-ph/0302326} {arXiv:astro-ph/0302326 [astro-ph]}
  \BibitemShut {NoStop}%
\bibitem [{\citenamefont {Weinberg}(2004{\natexlab{a}})}]{Weinberg:2004kr}%
  \BibitemOpen
  \bibfield  {author} {\bibinfo {author} {\bibfnamefont {S.}~\bibnamefont
  {Weinberg}},\ }\href {\doibase 10.1103/PhysRevD.70.043541} {\bibfield
  {journal} {\bibinfo  {journal} {Phys.Rev.}\ }\textbf {\bibinfo {volume}
  {D70}},\ \bibinfo {pages} {043541} (\bibinfo {year} {2004}{\natexlab{a}})},\
  \Eprint {http://arxiv.org/abs/astro-ph/0401313} {arXiv:astro-ph/0401313
  [astro-ph]} \BibitemShut {NoStop}%
\bibitem [{\citenamefont {Weinberg}(2004{\natexlab{b}})}]{Weinberg:2004kf}%
  \BibitemOpen
  \bibfield  {author} {\bibinfo {author} {\bibfnamefont {S.}~\bibnamefont
  {Weinberg}},\ }\href {\doibase 10.1103/PhysRevD.70.083522} {\bibfield
  {journal} {\bibinfo  {journal} {Phys.Rev.}\ }\textbf {\bibinfo {volume}
  {D70}},\ \bibinfo {pages} {083522} (\bibinfo {year} {2004}{\natexlab{b}})},\
  \Eprint {http://arxiv.org/abs/astro-ph/0405397} {arXiv:astro-ph/0405397
  [astro-ph]} \BibitemShut {NoStop}%
\bibitem [{\citenamefont {Arkani-Hamed}\ \emph {et~al.}(2004)\citenamefont
  {Arkani-Hamed}, \citenamefont {Cheng}, \citenamefont {Luty},\ and\
  \citenamefont {Mukohyama}}]{ArkaniHamed:2003uy}%
  \BibitemOpen
  \bibfield  {author} {\bibinfo {author} {\bibfnamefont {N.}~\bibnamefont
  {Arkani-Hamed}}, \bibinfo {author} {\bibfnamefont {H.-C.}\ \bibnamefont
  {Cheng}}, \bibinfo {author} {\bibfnamefont {M.~A.}\ \bibnamefont {Luty}}, \
  and\ \bibinfo {author} {\bibfnamefont {S.}~\bibnamefont {Mukohyama}},\ }\href
  {\doibase 10.1088/1126-6708/2004/05/074} {\bibfield  {journal} {\bibinfo
  {journal} {JHEP}\ }\textbf {\bibinfo {volume} {05}},\ \bibinfo {pages} {074}
  (\bibinfo {year} {2004})},\ \Eprint {http://arxiv.org/abs/hep-th/0312099}
  {arXiv:hep-th/0312099 [hep-th]} \BibitemShut {NoStop}%
\bibitem [{\citenamefont {Easther}\ \emph {et~al.}(2011)\citenamefont
  {Easther}, \citenamefont {Flauger},\ and\ \citenamefont
  {Gilmore}}]{Easther:2010mr}%
  \BibitemOpen
  \bibfield  {author} {\bibinfo {author} {\bibfnamefont {R.}~\bibnamefont
  {Easther}}, \bibinfo {author} {\bibfnamefont {R.}~\bibnamefont {Flauger}}, \
  and\ \bibinfo {author} {\bibfnamefont {J.~B.}\ \bibnamefont {Gilmore}},\
  }\href {\doibase 10.1088/1475-7516/2011/04/027} {\bibfield  {journal}
  {\bibinfo  {journal} {JCAP}\ }\textbf {\bibinfo {volume} {1104}},\ \bibinfo
  {pages} {027} (\bibinfo {year} {2011})},\ \Eprint
  {http://arxiv.org/abs/1003.3011} {arXiv:1003.3011 [astro-ph.CO]} \BibitemShut
  {NoStop}%
\bibitem [{\citenamefont {Armendariz-Picon}\ \emph {et~al.}(2008)\citenamefont
  {Armendariz-Picon}, \citenamefont {Trodden},\ and\ \citenamefont
  {West}}]{ArmendarizPicon:2007iv}%
  \BibitemOpen
  \bibfield  {author} {\bibinfo {author} {\bibfnamefont {C.}~\bibnamefont
  {Armendariz-Picon}}, \bibinfo {author} {\bibfnamefont {M.}~\bibnamefont
  {Trodden}}, \ and\ \bibinfo {author} {\bibfnamefont {E.~J.}\ \bibnamefont
  {West}},\ }\href {\doibase 10.1088/1475-7516/2008/04/036} {\bibfield
  {journal} {\bibinfo  {journal} {JCAP}\ }\textbf {\bibinfo {volume} {0804}},\
  \bibinfo {pages} {036} (\bibinfo {year} {2008})},\ \Eprint
  {http://arxiv.org/abs/0707.2177} {arXiv:0707.2177 [hep-ph]} \BibitemShut
  {NoStop}%
\bibitem [{\citenamefont {Lopez~Nacir}\ \emph {et~al.}(2012)\citenamefont
  {Lopez~Nacir}, \citenamefont {Porto}, \citenamefont {Senatore},\ and\
  \citenamefont {Zaldarriaga}}]{LopezNacir:2011kk}%
  \BibitemOpen
  \bibfield  {author} {\bibinfo {author} {\bibfnamefont {D.}~\bibnamefont
  {Lopez~Nacir}}, \bibinfo {author} {\bibfnamefont {R.~A.}\ \bibnamefont
  {Porto}}, \bibinfo {author} {\bibfnamefont {L.}~\bibnamefont {Senatore}}, \
  and\ \bibinfo {author} {\bibfnamefont {M.}~\bibnamefont {Zaldarriaga}},\
  }\href {\doibase 10.1007/JHEP01(2012)075} {\bibfield  {journal} {\bibinfo
  {journal} {JHEP}\ }\textbf {\bibinfo {volume} {1201}},\ \bibinfo {pages}
  {075} (\bibinfo {year} {2012})},\ \Eprint {http://arxiv.org/abs/1109.4192}
  {arXiv:1109.4192 [hep-th]} \BibitemShut {NoStop}%
\end{thebibliography}
\end{document}